\documentclass[journal]{vgtc}                
\ifpdf
  \pdfoutput=1\relax                   
  \usepackage{graphicx}                
  \DeclareGraphicsExtensions{.pdf,.png,.jpg,.jpeg} 
\else
  \usepackage{graphicx}                
  \DeclareGraphicsExtensions{.eps}     
\fi%

\graphicspath{{figures/}{pictures/}{images/}{./}} 

\usepackage{microtype}                 
\PassOptionsToPackage{warn}{textcomp}  
\usepackage{textcomp}                  
\usepackage{mathptmx}                  
\usepackage{times}                     
\usepackage{cite}                      
\usepackage{tabu}                      
\usepackage{booktabs}                  

\usepackage[dvipsnames]{xcolor}

\newcommand{\rev}[1]{\textcolor{black}{#1}}

\onlineid{0}

\vgtccategory{Research}
\vgtcpapertype{evaluation}

\title{The Influence of Visual Provenance Representations on Strategies in a Collaborative Hand-off Data Analysis Scenario}

\author{Jeremy E. Block, Shaghayegh Esmaeili, Eric D. Ragan, John R. Goodall, G. David Richardson}
\authorfooter{
\item
 Jeremy E. Block is with University of Florida. E-mail: j.block@ufl.edu.
\item
 Shaghayegh Esmaeili is with University of Florida. E-mail: esmaeili@ufl.edu.
\item
 Eric D. Ragan is with University of Florida. E-mail: eragan@ufl.edu.
 \item
 John R. Goodall is with Oak Ridge National Laboratory. Email: jgoodall@ornl.gov.
 \item
 G. David Richardson is with Oak Ridge National Laboratory. E-mail: richardsongd@ornl.gov.
}

\shortauthortitle{Block \MakeLowercase{\textit{et al.}}: The Influence of Visual Provenance Representations on Strategies in a Collaborative Hand-off Data Analysis Scenario}

\abstract{
Conducting data analysis tasks rarely occur in isolation.
Especially in intelligence analysis scenarios where different experts contribute knowledge to a shared understanding, members must communicate how insights develop to establish common ground among collaborators.
The use of provenance to communicate analytic sensemaking carries promise by describing the interactions and summarizing the steps taken to reach insights. 
Yet, no universal guidelines exist for communicating provenance in different settings.
Our work 
focuses on the presentation of provenance information and the resulting conclusions reached and strategies used by new analysts. 
In an open-ended, 30-minute, textual exploration scenario, we qualitatively compare how adding different types of provenance information (specifically data coverage and interaction history) affects analysts' confidence in conclusions developed, propensity to repeat work, filtering of data, identification of relevant information, and typical investigation strategies.
We see that data coverage (i.e. what was interacted with) provides provenance information without limiting individual investigation freedom.
On the other hand, while interaction history (i.e. when something was interacted with) does not significantly encourage more mimicry, it does take more time to comfortably understand, as represented by less confident conclusions and less relevant information gathering behaviors.
Our results \rev{contribute} empirical data towards understanding how provenance summarizations can influence analysis behaviors.
} 

\keywords{Analytic provenance, sensemaking, information transfer, visualization, workflow summarization, user studies}

\vgtcinsertpkg

\begin{document}


\firstsection{Introduction}

\maketitle
Exploratory analysis involves the process of gathering information, identifying patterns, and investigating hypotheses.
Due to the open-ended nature of exploratory analysis, there can be uncertainty in understanding the thought processes and the factors contributing to conclusions~\cite{battle_characterizing_2019,Ragan_Goodall_2014}.
With multiple ways of working through the data, individuals might arrive at different conclusions.
Often this work happens in collaborative sessions or team environments~\cite{madanagopal_analytic_2019,zhao_supporting_2018}, so communicating how a discovery is reached is critical to all parties maintaining understanding with each other.


By assisting with tracking the analysis history, software can help facilitate collaboration and hand-offs of information across shifts.
By tracking the \textit{analytic provenance}~\cite{North_Chang_Endert_Dou_May_Pike_Fink_2011,Ragan_Goodall_Tung_2015,battle_characterizing_2019} of an investigation,  other analysts can later review the history to reveal what information was considered---or not considered---and how connections in the data led to the development of hypotheses or conclusions.
While the potential value of provenance information is strong, core challenges remain with how to process provenance data and design effective representations to support easy human understanding.
As has been well documented in the visualization community, the representation of information can have dramatic effects on human interpretation of data~\cite{feng_hindsight_2017,valdez_2018_priming,dimara_2019_mitigating,Wall_Narechania_Coscia_Paden_Endert_2022}.
Furthermore, there is relatively limited empirical knowledge of how provenance information is used in hand-off scenarios where a second analyst continues an analysis with provenance records from a prior analyst~\cite{zhao_supporting_2018}.
And while it is expected that awareness of a previous analyst's thought process would influence the analysis strategies for a new analyst, there is a need to better understand \textit{how} the approach might be affected.
Additionally, our research studies how different forms of provenance representation might influence continuing analysis behaviors in collaborative hand-offs.



We conducted a user study on how people use provenance information when continuing an analysis after a previous analyst's partial progress.
The study is situated in the context of an intelligence analysis scenario with a text data set.
To explore different analyst responses to provenance representation, the study compares two types of summarized provenance representations (based on interaction history and data coverage) along with a control condition without explicit provenance information.
Through this study, we characterize patterns in participants' analysis findings, interaction behaviors, and approaches to using the provenance information in an information hand-off scenario.
\rev{We contrast the effects of two provenance representations to show how confidence in one's conclusion relates to the level of detail in handed-off data representations, describe interaction metrics for investigation behaviors, and confirm analysis strategies identified in prior work.} 

\section{Related Work}\label{sec:relatedwork}
Here we describe the nature of sensemaking, the uses for provenance information to represent how users understand problems, and the various techniques used to evaluate and correct for bias in analyst processes.

\subsection{Collaborative Sensemaking}

From the early work of decision theory~\cite{Tversky_Kahneman_1974,Bruckmaier_Krauss_Binder_Hilbert_Brunner_2021}, the goal was to understand how users arrived at a clear conclusion when working with ill-structured data. 
Researchers continued this work with the study of the \textit{sensemaking} process, which covers the human tendency to oscillate between foraging for new information and schematizing how it fits with what one already knows~\cite{pirolli_sensemaking_2005}.
While multiple definitions have evolved from this preliminary work, the general understanding is that sensemaking ``involves the ongoing retrospective development of plausible images that rationalize what people are doing''~\cite{weick_organizing_2005}.
Of interest to the work discussed in this paper is the \textit{Data-Frame} model proposed by Klein et al.~\cite{klein_data-frame_2007}.
In their work, they explain \textit{how} something is introduced can have a direct impact on how users frame and continue their analysis.
Intelligence analysis~\cite{endert_semantic_2012,bradel_multimodel_2014}, medical diagnosis~\cite{gonzalez_handoff_2018,murray_medknowts_2021}, and even humble internet research~\cite{nguyen_sensepath_2016,nguyen_sensemap_2016,rachatasumrit_forsense_2021} 
all share tasks where complex data requires thoughtful consideration and \rev{artifact} synthesis to arrive at and present a formal conclusion~\cite{sharma_2009_artifact}.
So, while common analytic settings clearly employ parts of the sensemaking process, they also define operations involving collaborative teaming or hierarchical units~\cite{madanagopal_analytic_2019}.
Collaborators can distribute workload, but they need to establish a common ground to understand the investigation status and contribute toward the goal~\cite{Robinson_2008}.

Even without the added complexity of coordination and knowledge sharing among collaborators, sensemaking tasks are often nonlinear by nature.
Insights and connections may be discovered independent of a strict method or procedural scaffolding, making it more challenging to systematically describe how concepts build on each other or explain the overall relationships.
Add to the combination that collaborating analysts are dealing with multiple layers of uncertainty and trust~\cite{sacha_role_2016}, and the situation becomes even more complicated.
        Among the many purposes targeted for provenance support in visual analysis applications~\cite{Ragan_Endert_Sanyal_Chen_2016}, aid for collaborative work is bolstered by helping people recall what they know, clearly communicate the steps taken, and making it possible to reproduce past work~\cite{sharma_2009_artifact}.
        As demonstrated by Mathisen et al.~\cite{Mathisen_Horak_Klokmose_Gronbak_Elmqvist_2019} in their description of the \textit{InsideInsights} tool, capturing a user's interactions as they worked and providing an interface for the addition of annotations, improved the collaboration and established common ground.
        This technique of enhancing interaction data with analyst-generated annotations is a common way to help maintain context for an analyst or different audience members~\cite{park_storyfacets_2021}.
        Still, often the process of annotating (e.g., writing notes, tagging information) distracts from the gathering of information because it requires users to synthesize their fuzzy concepts into specific terms that may or may not communicate their exact meaning upon later review~\cite{sharma_2009_artifact}.
        At the same time, the process of exploratory data analysis is inherently dynamic, leading to the requirement for plans to constantly change with the situation~\cite{Suchman_1987}.
        Without the transcription of accurate mental schemes, details can be forgotten, leading to false conclusions and inaccuracies~\cite{Ragan_Goodall_Tung_2015,Ragan_Goodall_2014}.
        By partnering with computers, human analysts can focus less on annotation tasks, like recording how they arrived at different concepts, and shift their attention toward directing the analysis and hypothesizing relationships between discovered ideas~\cite{chinchor_science_2009,endert_human_2014,rachatasumrit_forsense_2021}.
        In this study, we examine how the representation of a user's process impacts the sensemaking processes of new analysts. 

\subsection{Provenance and Visual Summarization}

Analysis tools that capture analytic provenance information aim to improve the understandability~\cite{bao_sharing_2013}, reproducibility~\cite{pasquier_if_2017}, and transparency~\cite{heer_graphical_2008,chung_vizcept_2010} of insight generation over time by providing the ``story'' of data exploration and interpretation\cite{park_storyfacets_2021}.
Yet, there are a variety of provenance types that serve different purposes depending on the context~\cite{Ragan_Endert_Sanyal_Chen_2016,herschel_survey_2017}.
Provenance can aid in the recall of past work, the verification of others' work, the recovery of past actions, the review and optimization for future analysis, the presentation of new information, and ultimately the communication of insights between collaborators~\cite{Ragan_Endert_Sanyal_Chen_2016}.

Visual provenance summarizations describe events that occur over time, and myriad approaches have been demonstrated for both algorithmic summarization and visual representation to ease human interpretation of the workflow, especially in collaborative settings.
There is a need for accurate techniques that compress temporal events that occur while matching an appropriate level of temporal granularity and summarization to best serve different audiences.
Some techniques focus on preserving the timing and order of events to allow for the review of specific analysis turning points (i.e., \textbf{History} representations)~\cite{dunne_graphtrail_2012,nguyen_sensepath_2016, zhao_supporting_2018} while others provide a high-level summary of topics reviewed and remove elements of timing completely to make it easy to see what has been explored and what needs further analysis (i.e., \textbf{Coverage} representations)~\cite{Wall_Narechania_Coscia_Paden_Endert_2022,sarvghad_visualizing_2017,feng_hindsight_2017}.
Provenance helps collaborators to maintain common ground as they work synchronously~\cite{chung_vizcept_2010}, or asynchronously~\cite{zhao_supporting_2018,xu_2018_chart}.
To assist in collaborator communication, many provenance visualizations focus on providing a reference to a user's interaction history or data actions using a timeline~\cite{zhao_supporting_2018} or branching trees~\cite{nguyen_sensemap_2016,dunne_graphtrail_2012,heer_graphical_2008}.
Others take a drastically different approach by removing the temporal aspect entirely and instead aim to describe \textit{what} data was explored rather than \textit{when} it was explored.
By ignoring time in the representation, viewers can focus on the context of data coverage and patterns (e.g.,~\cite{sarvghad_visualizing_2017,Wall_Narechania_Coscia_Paden_Endert_2022,xu_2018_chart,Zhao_Fan_Feng_2022}).
These techniques trade a higher level of summarization 
for less emphasis on analysis step replication and the specific actions taken to arrive at an analysis state.



The potential value of provenance summarization features in analysis tools is well justified by prior literature~\cite{heer_graphical_2008, oliveira_provenance_2018,Ragan_Endert_Sanyal_Chen_2016,xu2020survey,xu_2018_chart,Zhao_Fan_Feng_2022}.
However, it is less clear how differences in the way the provenance information is summarized and visualized can affect an analyst's process and decision-making.
Numerous studies from the visualization community have demonstrated that differences in visual representation or the addition of new information can influence user bias.
For example, Dimara et al.~\cite{dimara_2019_mitigating} have shown how allowing users to intentionally remove salient data (e.g., outliers) can lead to less bias and more rational decisions.
Also, Wall et al.~\cite{Wall_Narechania_Coscia_Paden_Endert_2022} have shown how providing users a summarization of what they have reviewed (i.e., an overview of their data coverage) can increase some user's awareness of unconscious biases, while potentially encouraging others to amplify their biases by intentionally focusing their analysis on specific areas or hypothesis too.

Considering the common aim of using provenance visualization to support collaboration among multiple analysts~\cite{chung_vizcept_2010,sarvghad_exploiting_2015}, and the impact that visualization can have on user interpretations\cite{Cho_Wesslen_Volkova_Ribarsky_Dou_2017}, there is a need to better understand the ways the availability and representation of provenance information from one analyst may influence another analyst's behaviors, biases, or conclusions.
In addressing this gap, our work draws from lessons learned from existing empirical studies on visual design, provenance, and users' analysis choices to further understand how to best provide provenance information to users.

\subsection{Empirical Studies of Visual Design Influences}
Within the visualization community, there is a long history of evaluating the effects of using visual tools on user performance which is consistent with provenance representations.
Of specific interest are the behavioral effects when making history information available.
For example, Zhao et al.~\cite{zhao_supporting_2018} focused on the various strategies participants used when examining the work of collaborators, suggesting that different strategies lead to different degrees of investigation completeness.
Similar work has evaluated the types of strategies users employ when completing sensemaking tasks~\cite{Kang_Gorg_Stasko_2009} and there are concerns that the strategies are influenced by the interface.

This becomes quite obvious when considering the types of interactions afforded to users in an interface. 
For example, tools like \textit{Hindsight} clearly describe recall provenance through scented widgets to make it clear what parts of the data individuals have already explored in more detail~\cite{feng_hindsight_2017}.
By lowering the opacity for parts of the data recently examined, unexplored areas were, therefore, more prominent, and users were influenced to extend their examination.
\rev{Xu et al.~\cite{xu_2018_chart}, in a more collaborative setting, similarly invited participants to review what data had been previously inspected and how it was visualized by prior analysts.
By spatially distributing others' attempts in a ``constellation,'' finding commonly used data and alternative areas of interest become more available to the user.}
This is a direct result of the interface and its influence on user interactions.
Consistent with this line of work, multiple studies have shown how cognitive factors (like the anchoring effect~\cite{cho_anchoring_2017}) can significantly impact users' insights and exploration in decision-making tasks~\cite{Cook_Smallman_2008}.

It is clear that representations influence analyst behaviors~\cite{sacha_role_2016}.
    One of the goals of analytic provenance research is to prevent the effects of belief perseverance~\cite{cho_anchoring_2017}, base rate biases~\cite{Narechania_Coscia_Wall_Endert_2021}, misuse of representative heuristics~\cite{AlKhars_Evangelopoulos_Pavur_Kulkarni_2019}, and resolving conflicting insights~\cite{li_resolving_2020}. 
    As an example, visualization techniques have tried to prevent selection bias by showing an overview of how the data has been filtered, 
    and data type comparisons to help users recognize when they may be examining a subset of data too closely or with too much emphasis~\cite{Borland_Wang_Zhang_Shrestha_Gotz_2020}. 
Similarly, by showing the work already complete and suggesting ways for the analysis to continue, SOMflow also helped analysts complete a more thorough investigation ~\cite{Sacha_Kraus_Bernard_Behrisch_Schreck_Asano_Keim_2018}. 
    These techniques rely on provenance information to help users recall what has been explored and potentially rectify their cognitive biases.
\rev{Applicable to this study, Sarvghad and Tory~\cite{sarvghad_exploiting_2015}, compared how coverage and timeline representations improved an analyst's accuracy, and the amount of data explored when analyzing structured numerical data. 
To extend their findings, we compare effects in textual data analysis and also inspect the strategies employed.}
    Although there is evidence that provenance representation influences user strategies and performance, there are outstanding questions about how different provenance representations compare\rev{, especially in textual data analysis}. 
    %



\section{Experiment} 

To understand the influence of provenance representations on continuing open-ended investigations, we conducted a between-subject experiment.
Participants were asked to complete an exploratory data analysis task started by a prior investigator with different representations of provenance information available.
We describe the key factors of interest and experimental design in the next section. 

\subsection{Motivation and Study Design}
Designs for summarizing analytic provenance can take a wide variety of forms.
For the focus of our study, we consider two common classes of provenance summaries as generalizations of designs found in the research literature.
Specifically, we distinguish provenance representations into either \textit{interaction history} or \textit{data coverage} summaries.
Both designs bring uniquely different benefits to analysts in practice.

Provenance summaries using an \textbf{interaction history} approach tend to provide a timeline of events or describe which data are processed over time.
        This type of provenance helps identify critical moments of failure, reproduce results, or provide additional data transparency, but it takes more time to review because there is often more data to make sense of.
Unfortunately, while computing systems are capable of capturing interaction events, this process can quickly bloom into large sequences that are challenging to summarize.
Many types of provenance visualization tools do not distill meaning from raw interaction logs, choosing instead to visually represent all interactions or analysis stages in interactive tools to uncover patterns or flows (e.g.,~\cite{chung_vizcept_2010,heer_graphical_2008}). 

In contrast to provenance designs emphasizing the temporal flow of the analysis, we describe a separate generalized class of provenance summary as \textbf{data coverage} designs, which typically represent an overview of \textit{what} data was explored instead of \textit{when} it was explored.
By compressing time, users can see what has been explored and what remains~\cite{sarvghad_exploiting_2015,sarvghad_visualizing_2017,Wall_Narechania_Coscia_Paden_Endert_2022,xu_2018_chart,Zhao_Fan_Feng_2022}.
These techniques provide a higher level of summarization, focusing on providing a sense of context but lack enough detail to clarify or recreate past work~\cite{borkin_2013_evaluation}.

While modern techniques still implement examples from both data coverage and interaction timelines~\cite{zhao_supporting_2018,Stitz_Gratzl_Piringer_Zichner_Streit_2019,Walchshofer_Hinterreiter_Xu_Stitz_Streit_2020}, prior work has seen greater emphasis on using machine learning to extract patterns and assist in the summarization of time~\cite{Guo_2016_case,Gove_2021_automatic,Shi_Xu_Sun_Shi_Cao_2021,Zhao_Fan_Feng_2022} and less emphasis on comparative studies investigating the implications of different provenance summaries with people.
Questions remain about how best to summarize interaction histories in digestible ways that help analysts by balancing content and cognitive load. 
Many works have shown that provenance representations influence the analytical behaviors of users~\cite{bao_sharing_2013,Cho_Wesslen_Volkova_Ribarsky_Dou_2017,feng_hindsight_2017,Goyal_Fussell_2016}, yet they do not directly compare the effects of \rev{the prototypical} provenance representations we discussed.
A more direct comparison between interaction history and data coverage would be beneficial since both techniques summarize and present the past in a digestible way and future automation techniques would benefit from guidance on selecting the appropriate level of detail for a user's task.




Therefore, we designed a between-subject experiment to study how individuals work with different forms of provenance information while completing a textual data investigation. 
\rev{We address the following research questions to help direct our analysis:}
\begin{enumerate}
    \item \rev{RQ1: How does the inclusion of analysis history or data coverage influence the \textbf{conclusions} reached by a secondary analyst?} 
    \item \rev{RQ2: How do secondary analyst \textbf{behaviors} differ when provided the analysis history or data coverage from a prior analyst?}
    \item \rev{RQ3: How does the inclusion of analysis history or data coverage influence the \textbf{types of strategies} a secondary analyst uses to solve the problem?}
\end{enumerate}

\rev{As a basis for the study, we used a browser-based, direct manipulation interface to display documents and record participant interactions.}
Since provenance representations are commonly used in collaborative data analysis scenarios, we simulated a hand-off scenario where users pick up and finish the analysis started by someone else.
In an online study, participants were asked to review a set of documents and describe any associations they were able to make.
Provenance representations informed by the same prior analyst allowed for comparing analyst conclusions and strategies. 
Based on prior work with provenance, we hypothesized two main effects on behavior in collaborative hand-off cases.
\rev{We expected history summaries to encourage the continuing analyst to engage in more verification of the prior analyst's progress due to the inclusion of a more complete record of the prior analyst's work (\textit{Hypothesis H1}).
And since data coverage summaries provide context about what has been explored at a glance, we expected users to quickly understand the prior analysis and explore other topics (\textit{Hypothesis H2}).}

With a combination of a think-aloud protocol~\cite{nielsen2002getting}, screen recordings, interaction logs, and semi-structured interviews, we sought to identify differences in participant conclusions reached and strategies used to better characterize the impact of provenance on collaborative analysis. 
  
\begin{figure}
    \centering
    \includegraphics[width=\linewidth]{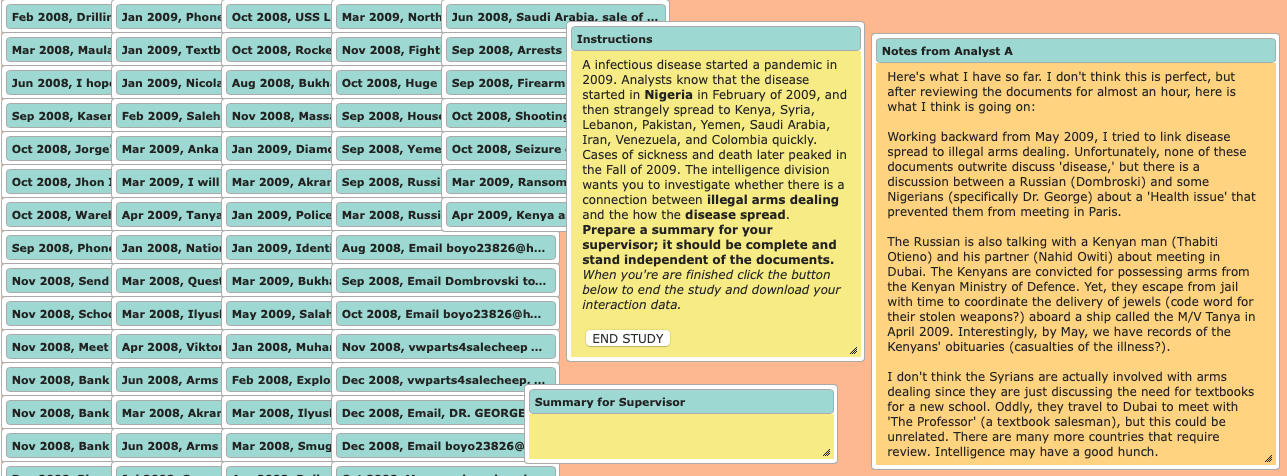}
    \caption{The textual analysis interface consisting of teal headers that open to reveal documents. Participants were given a summary from a prior analysis session. Some participants also received a representation of the prior analyst's provenance information as described in Figure \ref{fig:conditions}.}
  \label{fig:teaser}
\end{figure}

\subsection{Visual Analysis Task and Tool}\label{sec:task}

In collaborative data analysis tasks, users work together to uncover relationships and share results.
Frequently, these analysis tasks require users to communicate ill-structured and potentially relevant information for future analysts to recognize and use.
For our study of how a second analyst picks up after a prior analyst makes partial progress,
we used a synthetic intelligence analysis scenario based on an existing publicly available dataset from the first micro challenge of the 2010 VAST Challenge data series~\cite{grinstein_VAST_2010}.
\rev{This dataset, and others from the VAST Challenge, are commonly used as realistic proxies to simulate analysis tasks for visualization research (e.g.~\cite{zhao_supporting_2018,sarvghad_exploiting_2015}).} 
The data consists of fictional phone transcripts, email correspondence, forum posts, newspaper articles, and other intelligence reports \rev{about fictional} illegal arms traders.
\rev{Of the 103 documents in the dataset, only 16 contain information relevant to the solution we asked participants to complete.}

\rev{Within the tool, users could flexibly move and collapse documents as they explored, similar to other analysis workspace tools~\cite{Keel_2006,andrews_2010_space,Ragan_Goodall_Tung_2015}}.
Specifically, users were tasked with determining if illegal arms traders were responsible for the spread of a mysterious pandemic.
\rev{Participants could right-click to access a context menu.
From this menu, they could trigger a search event for the term under their cursor or type out their own query in a text field. 
Searches highlighted the set of document title bars that contained exact string matches to the queried text.} 
Critical to this experiment was identifying when and what kinds of information was revealed to users.
The tool would actively log user events (e.g., document opens, mouse enters and exits, and searches conducted) as they explored the dataset.
Each logged interaction was recorded with the time since the session began, the event type, the element's identifier, and other relevant information (position on screen, content of search, etc.).
All participants were given the same set of documents, instructions, a ``Summary for Supervisor'' field, and a note from the prior analyst (see figure ~\ref{fig:teaser}). 
The instructions further reiterated the scenario introduced to the participants and 
the ``Summary for Supervisor'' was a blank note where participants would type out their conclusions at the end of the session. 
\rev{We describe the provenance representations and their interactions in the next section.}

    \begin{figure}
        \centering
        \includegraphics[width=\linewidth]{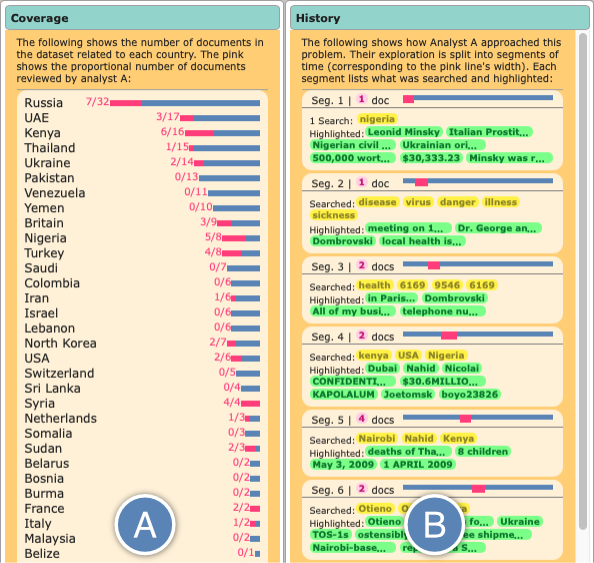}
        \caption{Participants belonged to one of the three conditions: Control (no view), Control + the Coverage panel (A), or Control + History panel (B). When clicked, the countries (A) or segments (B) would help filter the documents related to the associated data. 
        }
        \label{fig:conditions}
    \end{figure} 
    
\subsection{Conditions}\label{sec:conditions}
As described in Section \ref{sec:relatedwork}, \rev{we define} two general categories for provenance representations based on how they communicate time.
To further explore how data coverage and interaction histories \rev{impact investigation conclusions, behaviors, and strategies}, we examined prototypical representations of each.
Below, we describe the different provenance representations used in the experiment:

        To serve as a \textbf{Control}, all participants were given a textual description designed to mimic the series of conclusions \rev{or cursory annotations} a prior analyst may string together as they completed the same analysis scenario (i.e., the ``Notes from Analyst A'').
        This note served as an easy way to provide context and offer potential entry points for a participant's data exploration.
        \rev{Inspired by similar methodologies~\cite{sarvghad_exploiting_2015,zhao_supporting_2018}, t}he note was based on the interaction log of a researcher's pantomimed analysis that followed specific details through documents to arrive at a partially correct conclusion.
        Several statements were intentionally hedged to provide ample openness for where participants could begin.
        In the pantomimed interactions, 15 documents were opened, along with 31 searches and 53 highlights.
        Generally, the referenced documents and approach would uncover the majority of details required to solve the whole solution \rev{(i.e., 6/16 documents)}.
        \rev{These participants were not given additional panels to filter the dataset and had to rely on the search tool described earlier to find information and solve the scenario.}
        We used this \rev{same} interaction history to construct the \rev{other} provenance representations discussed below.

        Some participants were additionally given a \textbf{Coverage} representation (Figure \ref{fig:conditions} A) of the data explored by the prior analyst.
        This view showed the countries that received the most attention from the prior analyst as a list with miniature bars.
        Each country's bar displayed the ratio of documents explored by the prior analyst 
        and the bars were arranged in descending order by the country's mention frequency.
        The original dataset~\cite{grinstein_VAST_2010} did not have labels for the countries mentioned in a document, so these had to be hand-labeled by the researchers.
        Intending to simulate how a hypothetical tool would work, the researchers attributed the first city mentioned in a document to its corresponding country.
        \rev{These hand-labeled countries were also added to the preamble of each document, to help balance the conditions. This way the other conditions could also filter specific countries with the search tool.}
        \rev{The Coverage panel made these country-focused searches conveniently accessible by displaying them as a list.}
        By clicking a country \rev{from the list}, a \rev{``tool-use''} event would be logged, and the affiliated documents would be revealed---distinguishing those explored by the prior analyst and those not reviewed---by coloring the documents' title bar.
        \rev{Clicking a selected country again removed document coloring.}
        \rev{Participants could select a country to filter the documents and help direct their investigation.}

        Other participants were provided a \textbf{History} representation (Figure \ref{fig:conditions} B) that summarized the steps the prior analyst took to complete their task.
        This view gathered and displayed the searches and highlights of analyst A as different segments of the analysis.
        The interactions from the baseline log were augmented with a handful of additional highlight terms or relevant search terms to provide a bit more content for the users of the History panel.
        \rev{Participants could scan through the History panel to help get a feel for the general terms (searches) and evidence (highlights) the prior analyst cared about over time.}
        While manually segmented, the segmentation was done systematically.
        Segments were delineated based on a search event and contained the corresponding highlight events that took place between searches.
        In total, there were 11 segments, each labeled with the number of documents reviewed as well as a small timeline \rev{in the right corner of each segment} to visualize its duration and placement in the prior analysis.
        Much like the Coverage representation, when a participant clicks a segment, the corresponding documents from that segment would be revealed by changing the color of affiliated document titles. 
        Clicking a search or highlight term would run the interface's search command and simulate the results the prior analyst would have seen.
        Clicking a segment again removed the applied colors.
        \rev{Participants could select segments to filter the documents and help review aspects of the prior analysis.}

\subsection{Procedure} 

\rev{This research was approved by the organization's institutional review board (IRB).}
Participants joined a virtual meeting room and were asked to complete a demographic questionnaire capturing age, gender, academic program, and self-report measures for the ability to complete analysis tasks and their communication abilities.
The experimenter then explained the web application interface and its functions to participants using a set of slides via screen-share.
Within the tutorial slides, participants were introduced to the think-aloud protocol and then asked to demonstrate the technique with a short, irrelevant document. 
The researcher offered feedback and ways to help improve their think-aloud (e.g., they were asked to read \rev{aloud} and verbalize their plans for what they would do next).
\rev{Also, it is known that first impressions can have a large degree of influence over what people focus on~\cite{tetlock1983accountability}, so all participants were read the same starting scenario (see section \ref{sec:task}) to control for variations in wording.}
Participants were invited to ask questions \rev{about interactions} throughout the tutorial, and a researcher was present with them in the interface to answer interface function questions as they worked.
Participants were not told to explicitly ``strategize,'' but rather to ``choose what to read with intention'' because ``they would not have enough time to read everything.''

\rev{Starting a screen recording,} participants were given 30 minutes to complete the task and asked to think aloud~\cite{fonteyn_description_1993} as they read and contemplated what associations they saw with the expectation that reading aloud would encourage deeper reflection~\cite{Block_Ragan_2020}.
At 10, 20, 25, and at the end of 30 minutes, participants were warned about the time elapsed and reminded of their task: ``Try to identify what associations may exist and prepare your
summary for your supervisor.''
After the analysis concluded, a post-task interview \rev{(included in the session recording)} helped capture their mental model and opinions of the tools they used.
With their final analysis workspace still visible, Participants were asked a series of semi-structured interview questions to further specify their understanding.
These questions asked participants about the relationships they were aware of, identify retroactive strategies they used to arrive at their conclusions, their thoughts about the prior analyst's work, and the provenance views as applicable.
Critically, they were asked to make a judgment on the relationship of arms dealing with the Nigerian disease (referred to as their conclusion).

\subsection{Participants}
We recruited 41 undergraduate university students from an upper-level computer science course as participants in the study.
Participants were compensated with course credits.
Data from five participants were excluded from analysis due to technical problems, communication uncertainty because of language barriers, or misunderstanding of task instructions.
Of the remaining 36 participants, \rev{they were initially randomly distributed, before researchers assigned later participants to the smallest groups to finish with experimental groups of equal size (12 participants per condition).}
Fourteen participants (38\%) identified as female, and 1 (3\%) identified as non-binary/third-gender. 
All but 6 students (83\%) were completing a degree in computer science, computer engineering, or software engineering. 

The majority of participants were between 18 and 24 years of age (77\%), 5 participants (14\%) were within the 25--34 age category and 3 participants (8\%) were older than 35.
We asked participants to self-report their ability to complete data analysis on a scale from 1--10; an average reported score of 3.14 signified general inexperience in completing analysis tasks, indicating the participant sample might be considered analogous to novice analysts with no or limited experience.

\newcommand*{\analyst}{\includegraphics[scale=2]{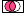}}
\newcommand*{\user}{\includegraphics[scale=2]{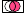}}
\newcommand*{\overlap}{\includegraphics[scale=2]{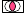}}
\newcommand*{\indpendence}{\includegraphics[scale=2]{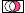}}

\section{Results}\label{sec:results}

In this section, we present the results and insights drawn from our quantitative and qualitative data analysis.
How people find and interact with the data can directly impact how they arrive at their conclusions and what conclusions they can make.
To understand the ways people use provenance representations, we looked at participants' analysis behaviors who pick up from a prior analyst’s progress.
We studied behaviors captured through video recordings, think-aloud comments, interaction logs, and post-study interview responses.

\subsection{User Findings and Confidence}
We studied whether the availability of the provenance views affected the participants’ findings and final conclusions \rev{(RQ1)} to look for implications of early bias influencing analysts’ ability to correct their preconceived expectations.
\rev{A single author scored each participant's written conclusions using a four-level rubric, according to four }factors, including the
\textbf{accuracy} of their reported findings; 
the \textbf{recognition of errors} made in the prior analysis; 
the \textbf{number of findings and amount of detail} provided; 
and the \textbf{depth of relationships or connections} among entities and events in the data. 

These qualities were a set of features we expected differences in based on provenance representations, yet the scores varied greatly. 
\rev{Due especially to the open-ended nature of user-directed analysis, the diversity of user aptitude with analysis and the various ways participants could write their findings, }our analysis did not find systematic differences in participants’ written conclusions caused by the provenance conditions.
\rev{For example, when summarizing key findings, some participants formatted a formal report in prose, while others opted for a series of evidentiary bullet points.
Often the written conclusion left out parts of the analysis and therefore was not a fair representation of the area's a participant explored.
For a similar reason, we do not report on participant encounters with the 16 solution-relevant documents, because we wanted to understand what information ``stuck'' and was reported in their concluding thoughts.}
Due to the range of individual differences among participants and personal styles for reporting, the quality and completeness of written conclusions varied greatly.
\rev{We did not find meaningful differences in analyst written conclusions.}
Therefore, we prioritized the analysis of data from the personalized post-study interview questions as a basis for identifying differences in participant conclusions and analysis behavior.

As part of this analysis, we assessed participants' confidence in \rev{the post-task interview when describing their final answer}.
High confidence implies that the user has convinced themselves of a specific relationship, and we want to see if that behavior varied systematically with the experimental conditions.
\rev{One author coded participants' interview responses to the question regarding their final answer about the possible existence of a relationship in the data}
(i.e., the main investigation goal in the analysis scenario).
\rev{With the support from a second author to review the handful of edge cases, we separated participant responses }into two categories (\textbf{high} and \textbf{low confidence}) based on the number of hedge statements in their verbal conclusion.
To do this, \rev{in the interview, we asked participants if they identified a relationship between the concepts they were investigating.}
Participants who stated clearly that there was or was not a relationship were designated as \textit{high} confidence (55\%).
\rev{Alternatively, those who suggested only a potential relationship, a need for additional time to review the data, or uttered more than two hedged phrases (e.g., ``there might be...,'' ``I guess...,'' ``I think...,'', etc.), were placed in the \textit{low} confidence category (45\%).} 
\begin{figure}
    \centering
    \includegraphics[width=\linewidth]{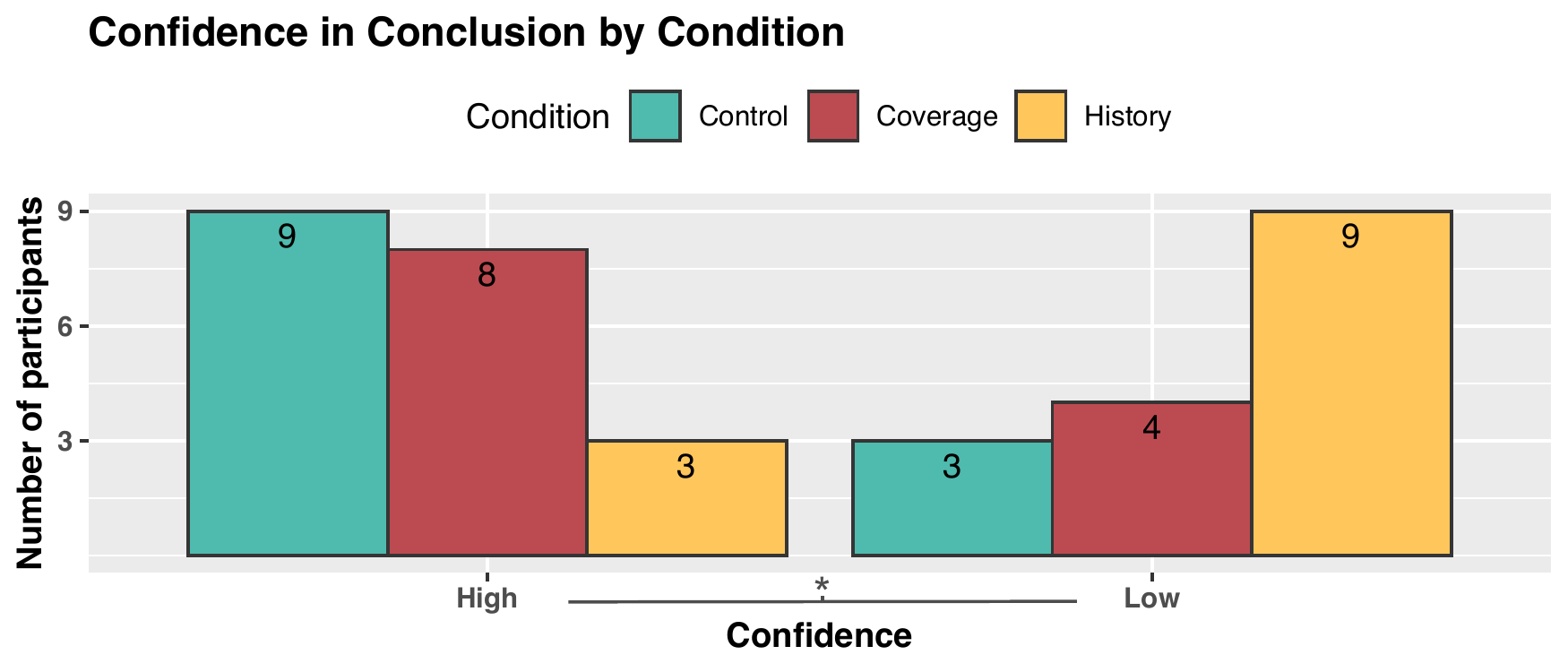}
    \caption{The distribution of participants' confidence in their conclusions after their analysis. While the condition appears to have an influence on the confidence of a participant, post hoc comparisons cannot determine which conditions lead to different confidences.
    }
    \label{fig:conclusion_confidence}
\end{figure}

Upon first inspection, Figure \ref{fig:conclusion_confidence} shows a strong difference in participant confidence.
We see a main effect with Fisher's exact test ($p < 0.05 $) by condition.
Yet, post hoc comparisons with Bonferroni correction fail to identify the specific differences between conditions.
While we cannot state one condition varied significantly from the others, those with a Coverage representation tend to exhibit conclusions more similar to the Control than those with the History representation.
This is to say that the majority of conclusion qualities are not as dramatically impacted by the representation of provenance information.
We do see some impact on participants' confidence when they verbally explain their conclusion, but without significant interaction effects, we instead turn our attention to the behaviors exhibited by participants to better explain how their analysis differed.

\subsection{Quantitative Indicators of Behaviors }
The way individuals interact with an interface indicates how they make sense of the data.
\rev{With the variety of user activities, we choose to use a quantitative indicator to more clearly describe behavior patterns in participant approaches (RQ2).}
As captured in their interaction logs, we turn to the various analysis events and actions participants ran in the interface as a quantitative proxy for their understanding and to tease out the influence of provenance representation on user analysis behaviors. 
We look at three key representations of interest: the degree of \textit{similarity}, and \textit{difference} to the prior analyst's review, and the \textit{rate of filtering} for specific information. 
\label{sec:similarity}
Of concern to the inclusion of provenance information is the influence on how much repetition is in subsequent analyses, or the amount of similarity to the prior analyst. 
While verification can be beneficial when auditing the veracity of a result, in most cases repeated work is not encouraged.
We wanted to compare the analysis of each participant to the referenced prior analysis to understand if the provided provenance representation influenced how they addressed the problem of ``continuing the analysis.''
If participants chose to look into the same concepts as the prior analyst, how similar was their review (\textit{overlap}), and if they looked at different things, how unique was their investigation (\textit{independence})?
To quantify the amount of overlap and independence in participants’ behaviors we looked at which documents in a participant's interaction history were opened. 
\rev{Since critical information in the underlying dataset is separated into individual documents, we can determine how similar an investigation was to another by examining the set of documents opened.}
Both provenance representations were based on the same set of 15 documents, and we calculated two \rev{separate, but similar} ratios (overlap and independence) from the sets of documents participants reviewed.
We calculated a participant's \textbf{overlap ratio} by considering the intersection between the set of documents a participant reviewed that was also reviewed by the prior analyst and divided by the number of documents the prior analyst reviewed ($\frac{\overlap}{\analyst} = \frac{\{User\} \cap \{Analyst\}}{\{Analyst\}}$).
An overview ratio of 1.0 implies that a participant saw all of the documents that analyst A reviewed.

The \textbf{independence ratio} captures the proportion of documents a participant reviewed that were different from the set reviewed by the prior analyst ($\frac{\indpendence}{\user} = \frac{\{User\}-\{Analyst\}}{\{User\}}$).
An independence ratio of 1.0 implies that a participant only reviewed documents that the prior analyst did not.
Although these ratios examine similar participant behavior properties, they are not inverse because they compare different document sets.

\begin{figure}[t]
    \centering
    \includegraphics[width=\linewidth]{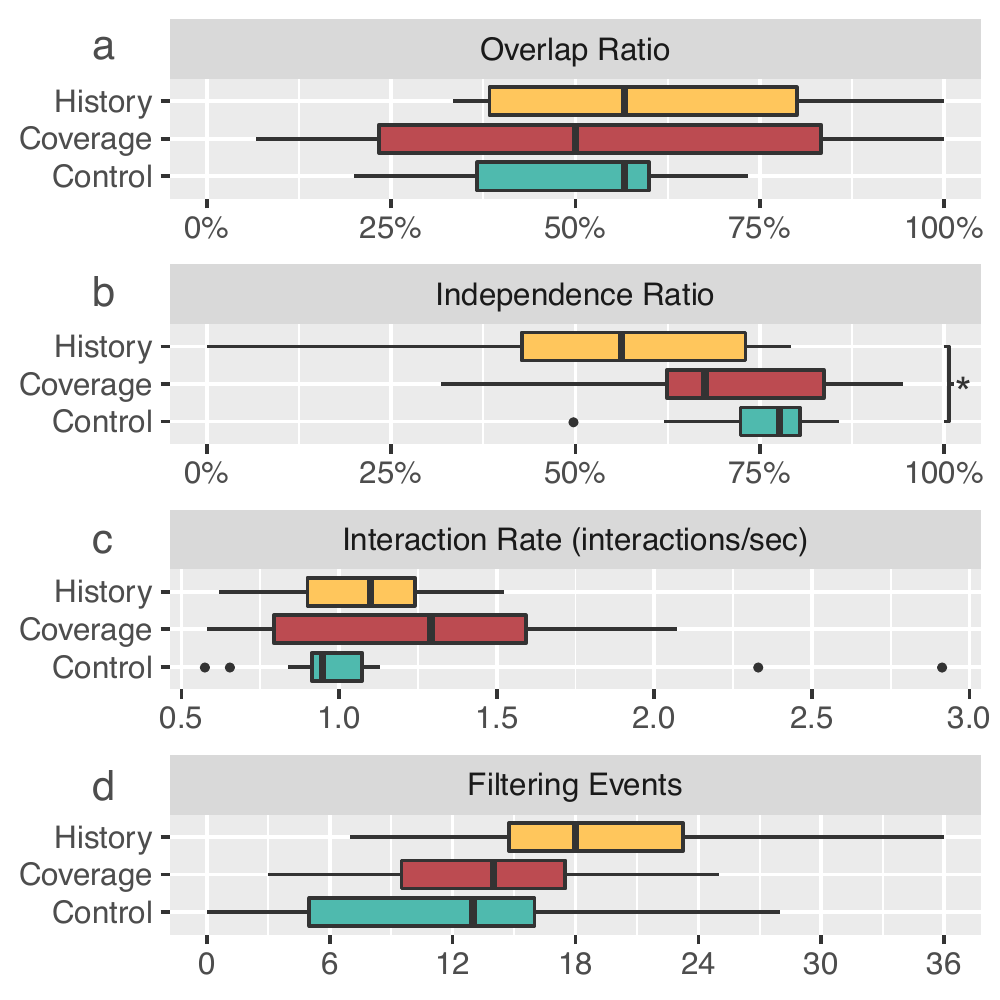}
    \caption{
    A user's interaction history is used to calculate the following.
    Both \textit{Overlap Ratio} and \textit{Independence Ratio} refer to the set of documents each participant opened and compares them to the prior analyst (see Section \ref{sec:similarity}
    ).
    \textit{Filtering Events} refers to interactions that help users find documents and is the combination of clicks in a provenance representation combined with search events.
    \textit{Interaction rate} is calculated as the ratio of total interactions over the length of analysis (i.e., interactions/sec).
    }
    \label{fig:metrics}
\end{figure}

When we compare the degree of overlap among participants (Figure \ref{fig:metrics}a), we do not see a strong difference between the conditions, implying that the amount of overlap a participant may have with a prior analysis has more to do with individual differences in approach and investigation intentions.
But, some expected trends are visible.
For example, those in the Control condition straddle the center ($\sim$50\%) as though they were unaware of which documents were reviewed by the prior analyst. 
We also see more spread in the Coverage condition, likely because they could filter documents and choose to follow or avoid the prior analysis.
Finally, the History condition has the highest median overlap ratio likely since their representation emphasized how the prior analyst worked through documents.

We see a much different behavior among participants’ ratios when we examine how independent their analysis was from the prior analyst (Figure \ref{fig:metrics}b).
A Kruskal-Wallis test revealed a significant difference in independence ratios $H(2) = 7.01, p<0.05$.
For those in the Control condition, more than half of the documents they reviewed had not been explored by the prior analyst.
The result can be explained since they had no idea which documents were specifically reviewed by the prior analyst, and opened many more documents on average.
With more documents opened, the likelihood of a document belonging to the set from the prior analyst goes down, leading to \rev{a higher independence ratio because there were many more documents that were not reviewed by the analyst.}
We also see a long tail for those in the History condition. 
The pairwise post-hoc Dunn test with Bonferroni adjustments showed to be only significant for the History and the Control ($p < 0.05$) suggesting that they generally spent their session reviewing mostly the same documents as the prior analyst.
They also looked at fewer documents overall. 
With fewer documents reviewed, and an emphasis on documents opened by the prior analyst, these participants were more likely to maintain lower Independence ratios.
Overall, we see that while the degree of overlap is not significant, there appear to be some differences in the diversity of documents participants are exposed to when given various provenance representations.
This is to say that the affordances provided by different tools can influence which information participants review.

\begin{figure}[ht]
    \centering
    \includegraphics[width=\linewidth]{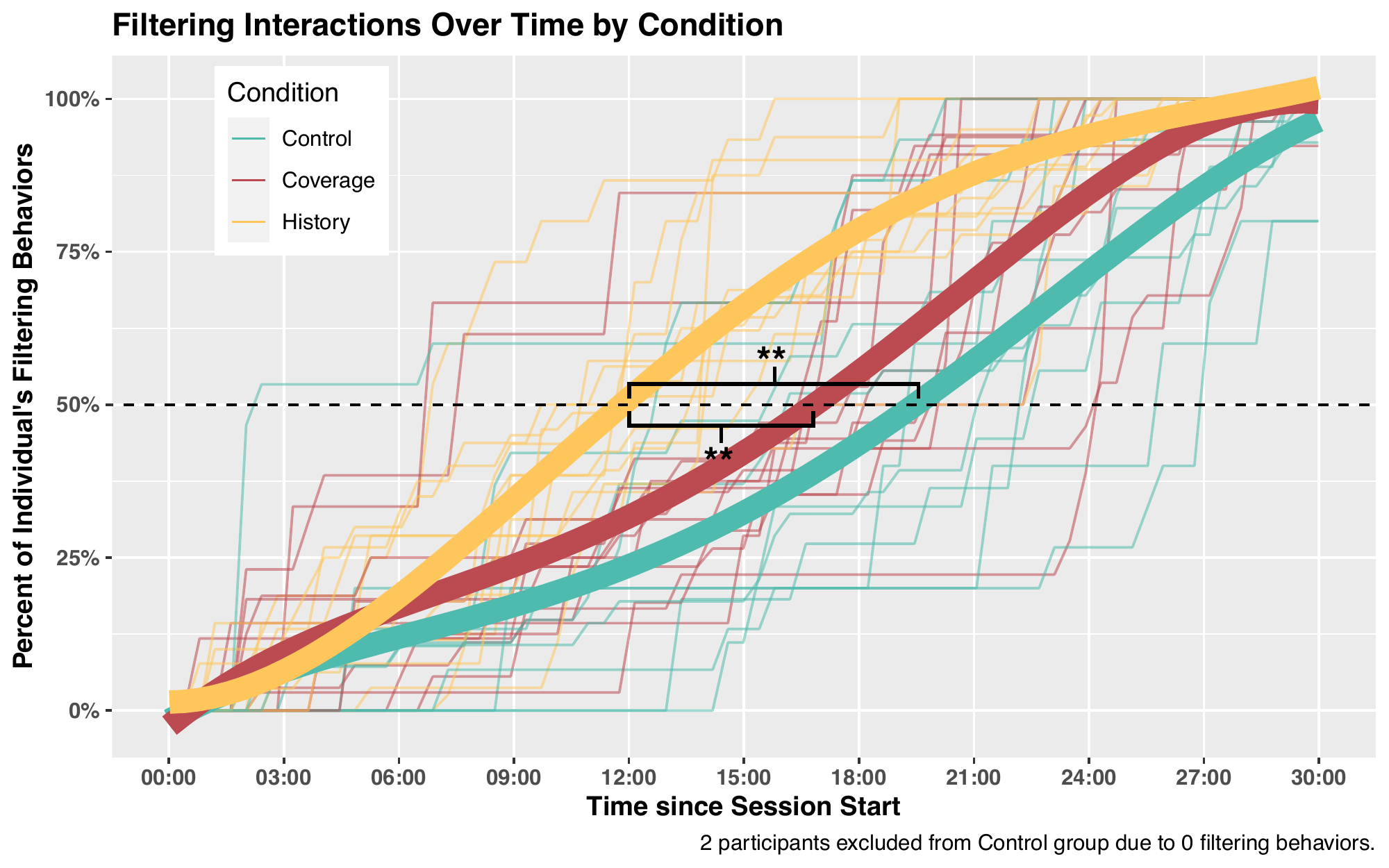}
    \caption{
    Trends in percent of a filtering interactions completed over time.
    The thicker lines represent a four-factor polynomial of trends by condition.
    Although the History condition takes some time to get started, they conduct more filtering earlier before tapering off later, whereas the Coverage and Control conditions used more filtering later in time and some continued to filter right up to the end. 
    }
    \label{fig:tool-use}
\end{figure}


Another aspect of our analysis relates to how participants direct their investigation. 
We looked at the total number of interactions and divided it by the length of a participant's session to determine an average rate of interaction.
The interaction rate (see Figure \ref{fig:metrics}c) can serve as a proxy for the level of control a participant has over the interface and their investigation.
Participants with exceptionally low interaction rates may be taking a long time to review documents or working very methodologically, whereas exceptionally high interaction rates may imply they have opened numerous documents at once or are not spending enough time understanding each document. 
For example, the Control condition opened the most documents (41.5 documents) on median, while maintaining the slowest median interaction rate (0.95 interactions/sec).
This contradictory phenomenon may suggest that those in the Control condition opened many more documents on median in an attempt to understand the data, but worked through the documents and the task slowly.
Yet, uncovering more descriptive interpretations would require the review of additional metrics. 

For example, we can turn to the frequency of filtering events to understand how participants search and reduce the data space (See Figure \ref{fig:metrics}d).
While the differences are not significant, there are more median filtering events completed (18.0) by those in the History condition while they also look at the fewest documents (28.5) on average. 
This is to suggest that those in the History condition were more likely to be methodological and consistent since they were not exposed to as many documents and spent more of their session filtering and refining their investigation criteria.
On the other hand, those in the Coverage condition appear to have overall investigation behaviors most similar to the Control, but also have the fastest interaction rate.
Those with the Coverage representation may have allowed participants to work faster and independently.

To further describe the ways participants were exposed to information, we examined the timing of users’ searches and provenance representation usage.
\rev{In this case, we define the term \textbf{filtering} to describe the sum of searches and provenance panel usage per participant.
Although those in the Control condition did not have access to a provenance representation, they relied more heavily on search to find information. 
This is to say that search and panel usage both helped to reduce the search space in the dataset and their combination provides a more universal comparison.}
We examine when these behaviors occur to understand when users are selecting and directing their investigations.
Since each participant filtered the data their own number of times, we calculate a filter percentage over time.
\rev{All but 2 participants eventually reached 100\% of their filtering events within their 30-minute session, but ultimately, we draw attention to the rate of change.}
In Figure \ref{fig:tool-use}, there is lots of overlap in when participants are completing their filtering behaviors.
To clarify the trends, we apply a 4-factor polynomial since it gave the highest $r^2$ coefficient (0.745) and helps characterize the behaviors observed.
As evidenced in the modeled regression, participants from the History condition complete about 50\% of their filtering \rev{events} by about 12 minutes, while it takes 17 or almost 19 minutes for the Coverage and Control conditions to conduct half of their \rev{filtering events} respectfully.
We counted the number of participants who had completed at least half of their filtering events by the halfway point in the investigation (15 minutes) as a proxy for how active participants were selecting data to review.
Because the provenance representations provided a more convenient (i.e., a single click) way to filter the data, 
we see that both the History and Coverage condition completed the majority of their filter interactions before those in the Control condition.
A Fisher's exact test confirmed that there were differences in the rate of filtering among the different conditions ($p < 0.001$).
Further post hoc pairwise Fisher's exact tests with Bonferroni corrections confirm that the amount of filtering completed by those in the History condition differed from the Control ($p = 0.001$) and Coverage conditions ($p < 0.001$), but no difference was detected between the Coverage and Control conditions at 15 minutes.
More specifically, those in the History condition complete the majority of their filtering behaviors before the other conditions (and in the first half of the session). 

\begin{figure}[t]
    \centering
    \includegraphics[width=\linewidth]{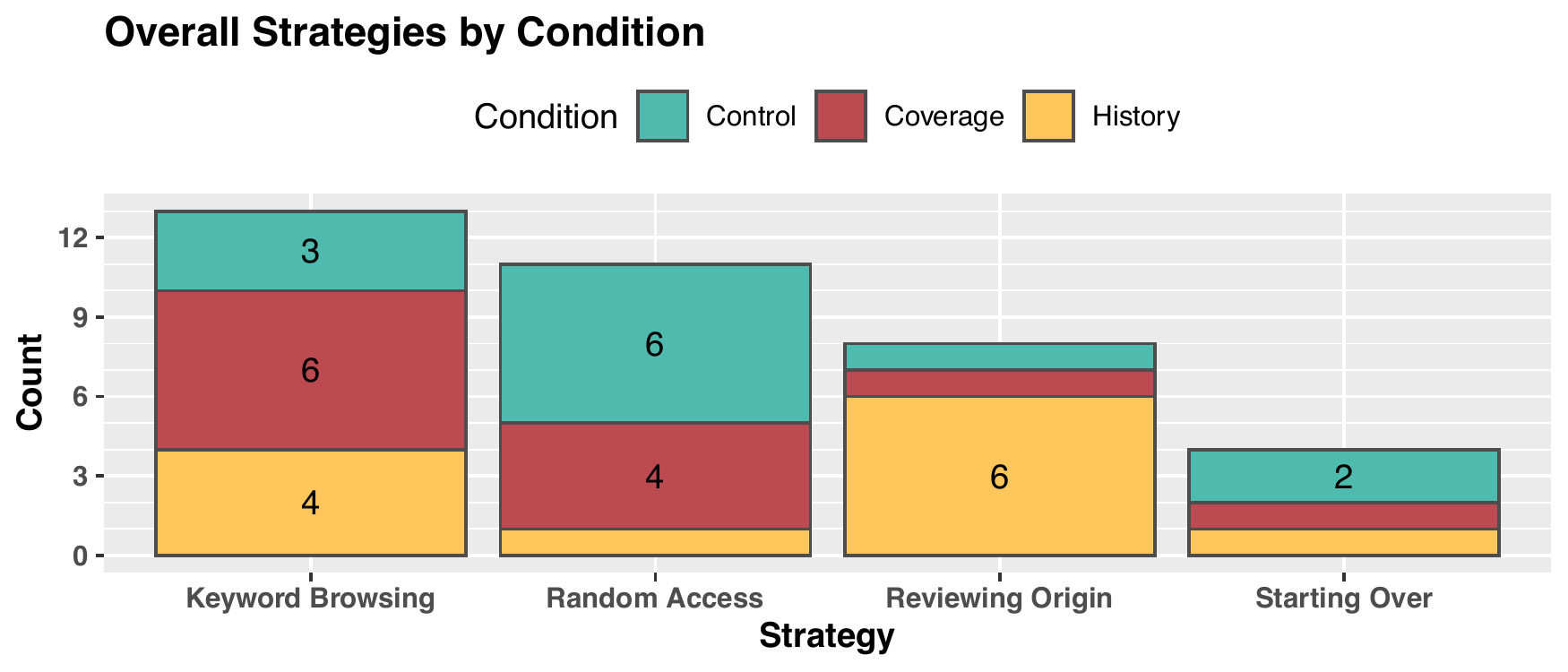}
    \caption{General strategies used by participants when continuing analysis.
    }
    \label{fig:strategies}
\end{figure}

We believe this difference is due to the interaction pattern required for the use of the History representation.
Since information about which documents were reviewed is only accessible after clicking a segment, participants from the History condition would commonly click through multiple segments at once in pursuit of a subset of documents to review instead of intentionally selecting a segment to review or independently typing out their own search query (discussed more in Section \ref{fig:strategies}).
This pattern of clicking through each segment in time, or scanning the resulting documents to find something previously reviewed likely led to an increase in filtering events overall (median 18.0 events) and also a higher frequency of events earlier in a session while participants were still gathering information.
On the other hand, it appears that the addition of Coverage information for participants did not significantly change when participants would be selecting/filtering data (as compared to the Control).
Ultimately, we see some quantitative differences in participant interactions, including how little content those in the History condition were able to review, and how frequently they were filtering the dataset.
We now take these quantitative differences and construct some qualitative definitions for user strategies.

\subsection{Qualitative Analysis of Strategies}
\rev{The strategies employed by participants during an open-ended evaluation likely depends on the kinds of information they are presented with (RQ3).}
Comparing and analyzing the strategies users take when approaching their exploration can shed light on how provenance information is used when investigating the relationships of various data.
In alignment with the work on situated planning~\cite{Suchman_1987}, of the participants who verbalized a preparatory plan, most were vague and often only consisted of 2--3 steps.
These early plans were interesting to us as we wanted to see if the availability of provenance information would influence how plans were made and adapted \rev{(see RQ3)}. 
With analysis plans being constantly adjusted and renegotiated as new information is acquired, we ultimately simplified our analysis by focusing on initial plans and the actions observed in the first 10 minutes of activity.

\rev{From participants' think-aloud and retrospective interviews, we tried to reconstruct participants' intentions as they began their analysis.
To do this, one author analyzed the data by conducting two rounds of open coding to establish a set of emergent features from user actions and analysis foci.}
\rev{In a similar data analysis task, Zhao et al.~\cite{zhao_supporting_2018} defined a set of analysis \textbf{strategies} derived from their own qualitative coding.
Their work compared the strategies participants used when constructing knowledge graphs with or without an interactive state timeline.
Our work differs in that we compare different provenance representations, while they only evaluate differences with or without provenance features.
While their study offered participants different data analysis affordances, we borrow similar qualitative analysis steps in our work as well.
From their identified strategy definitions, we cross-referenced our most common tags and refined a set of similar analysis strategies.
In contrast, ours are more focused on the amount of similarity to the original analyst's investigation as well as our users' commitment to their investigation plan.}
We clarify and define our categories below.

\begin{itemize}
\item \textbf{Keyword Browsing} - These participants set a plan for how they wanted to understand the data before they began or shortly after reviewing the note provided by the analysts. They made an intentional plan, maintained attention to the planned areas of interest, and frequently hypothesized different relationships. They had a medium amount of overlap with the prior analyst and tended to have a higher independence score since they were trying to extend the analysis and were more likely to have a more active role in the investigation.
\item \textbf{Random Access} - 	These participants had less structure to their investigations---often working without stating how they wanted to systematically approach the problem---and spending the majority of their time gathering information instead of synthesizing  hypotheses. These participants bounced around the dataset with about equal amounts of overlap and independence from the prior analyst.
\item \textbf{Reviewing Origin} - These participants expressed a plan to verify the work from the prior analyst. Often this was motivated by a lack of trust in the conclusions made in the analyst's summary or started verifying the prior analyst's work and ran out of time for their own investigation. Therefore these participants have noticeably lower independence scores and higher amounts of overlap.
\item \textbf{Starting Over} - These participants explicitly stated that they wanted to work independent of any influence from the prior analyst. They were worried explicitly about bias or interested in comparing their own understanding with the work by the prior analyst later. These participants intentionally closed the guidance from the prior analyst and waited till at least 15 minutes into the task before they reviewed the analyst's summary or the provenance panels.
\end{itemize}



\begin{figure}[tb]
    \centering
\includegraphics[width=\linewidth]{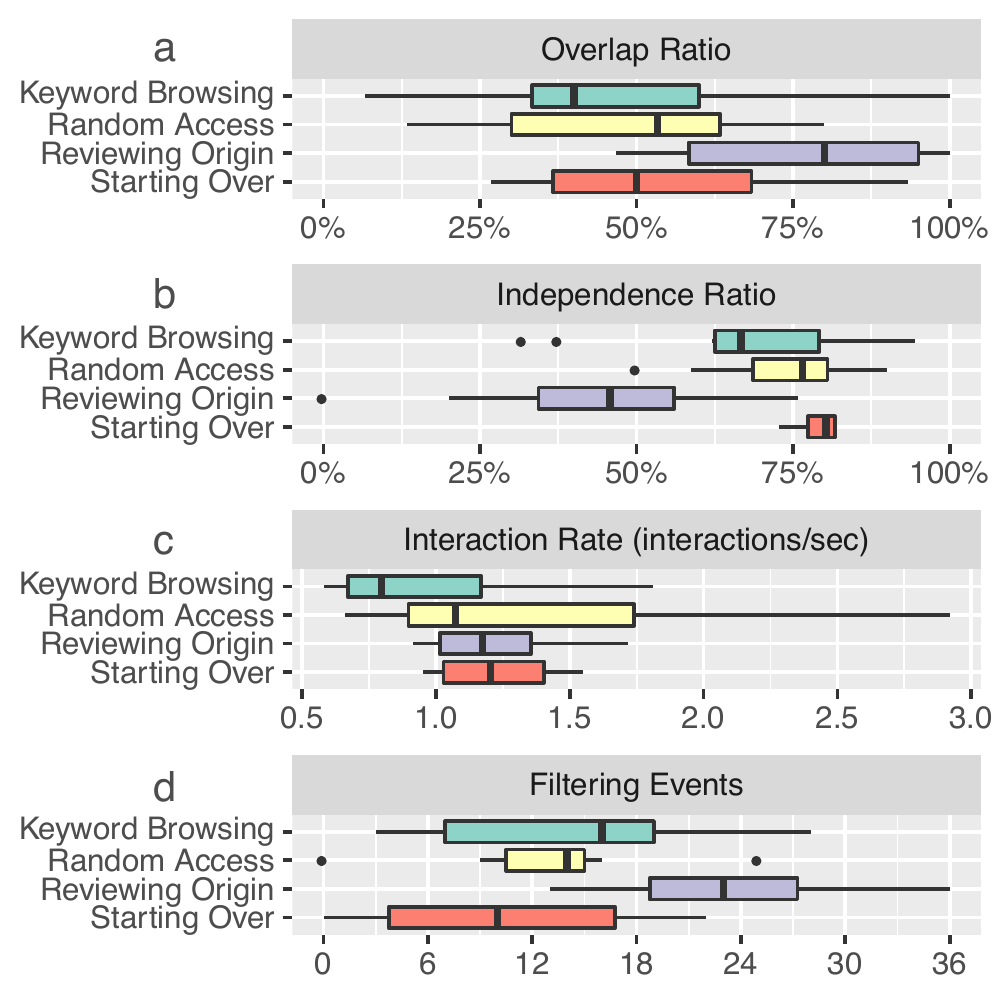}
    \caption{There are some interaction patterns among the participants in various strategy types. 
    Due to the limited participants in the \textit{starting over} (n=4), we do not compute statistical differences and instead focus on visual analysis and descriptive statistics.
    For details on the factors described, see Figure \ref{fig:metrics}.
    }
    \label{fig:strat-metrics}
\end{figure}

As reported in Figure \ref{fig:strat-metrics}, there are some interesting differences among users and their strategies.
With only a handful of participants in the \textit{starting over} group (n=4), we rely on descriptive statistics and visual analysis to describe the groups and their differences instead of standard statistical methods.
To begin, all groups appear to open the same number of documents, except for those in the \textit{starting over} group (45.5 document open events).
Those in the \textit{starting over} group also had the highest median interaction rate (1.20 interactions per second).
Much of this is likely due to the small number of participants in the group, but the way most of these participants worked was often without a stated plan and typically focused on gathering information generally. 
Similar behaviors were seen in the \textit{random access} group.
Without a plan, these participants also had high median interaction rates (1.07 interactions per second) and the highest interquartile range (0.90--1.74 interactions per second).
This can be explained by the lack of intention and purposeful activity these participants appeared to execute during the analysis and reduced specificity of what they were looking for. 
Frequently, these participants had bursts of activity where they would be gathering information slowly before they would find a recognizable piece of information and open multiple documents they already reviewed to find interrelationships and similarities.
Without a strong plan set at the beginning and generally less filtering to help find the information of interest, these participants were most likely to require more review of the data.

The other two categories had more explicit intentions for what they wanted to review.
Those in the \textit{reviewing origin} group completed the most filtering events on median (23.0 events). 
Likely due to the high ratio of participants from the History condition, this characteristically high number of filtering events is a direct result of using the History panel to review various segments of time; users had to trigger a filtering event each time they wanted to examine the documents reviewed in each segment of the prior analysis.
These participants also maintain a high degree of overlap with the prior analyst (80\%) and the lowest median independence (45\%).
On the other hand, those in the \textit{keyword browsing} group also had a stated plan and had a much different degree of overlap with the prior analyst (only 40\%).
Most of these participants intended to explore key aspects of the data described by the prior analyst and tended to open more documents the prior analyst had not touched. 
We noticed that those who used a keyword browsing strategy generally uncovered a more complete and accurate picture of the dataset.
We believe this is because they independently verified aspects of interest and followed keywords instead of the work by the prior analyst. 

Peculiarly, while there is not a significant difference, the distribution of strategies almost appear to follow with the three conditions(see Figure \ref{fig:strategies}), where \textit{keyword browsing} was mostly associated with those in the Coverage condition, \textit{random access} was mostly members from the Control condition and those in the \textit{reviewing origin} group were from the History condition.
Yet, without strong evidence that provenance representations influence the type of strategy participants employ when planning for their analysis, we cannot conclude that these strategies are determined by the affordances in the interface.

\section{Discussion}

Our work studies how providing provenance summaries can influence future investigators who pick up a partial analysis from a prior analyst.
We see evidence that: (1) listing interaction histories can result in a type of secondary sensemaking task for trying to understand the earlier analysis, (2) individual differences introduce a large amount of variation in how people choose to utilize provenance summaries, and (3) the types of strategies exhibited by the continuing analysts show some similarities with analytic strategies predicted by prior work.

    
    


    Considering the possible strategies participants adopted, we expected those who reviewed the prior analyst's progress would first start at the beginning of the prior analyst's work and then select relevant information to take into their own analysis and establish their own understanding.
    Yet, the observed results show the opposite trend.
    For example, the majority of participants in the \textit{reviewing origin} group were also from the \textit{History} condition.
    Interestingly, our interviews and observations of participants in the \textit{History} condition found they often felt overwhelmed with the extra available information, and we saw the majority of the \textit{History} condition had low confidence in their conclusions (see Figure \ref{fig:conclusion_confidence}).
    This corresponds with the frequent triggering and review interaction pattern required to see the documents reviewed by the prior analyst.
    This is also evidenced by the group's more frequent filtering behavior prior to 15 minutes, which may be due to the amount of information they were tasked to review in the limited time.
    We find \rev{supporting} evidence for $H1$, as it appears to take more time to construct an accurate mental model because participants were exploring the events in the data but also understanding how the prior analyst approached the problem.
     The broader implication is that
    having History information may exaggerate the task's difficulty by making more content accessible for review and not summarizing enough to serve users hoping to pick up a prior analysis. 
    On the other hand, providing Coverage information to users does not feel exceptionally beneficial beyond what was provided by the Control.
    While a greater proportion of participants with a Coverage representation may maintain more \textit{keyword browsing} strategies, there are no significant deviations from the control, thus rejecting $H2$.

    In open-ended analysis tasks, there are no clear paths that lead to a solution.
    Among the various metrics collected in our study to characterize analysis, we see the breadth of  approaches through the large degree of variation and spread in the data. 
    While some significant differences among conditions emerge (i.e., for the degree of investigation independence), many cross-condition differences may be hidden by the high variability in individual preferences and the way participants adapted their approaches in situ.
    Though the experiment provided the same instructions for the analysis tasks for all participants, we clearly found four unique types of approaches emerge.
    While a trend toward a specific strategy appears to align with the conditions (e.g., in Figure \ref{fig:strategies} we see 6 participants from each condition associated with different strategies), due to lack of significance we cannot conclude a direct relationship between provenance usage and strategy employed.
    A participant's choice to verify the prior work (i.e., \textit{reviewing origin}) or follow their own set of keywords (i.e., \textit{keyword browsing}) is likely a result of their personal experience completing analysis tasks or interest in referencing provenance information as well as other factors, and not due to how provenance information is provided.
    Expanded knowledge on the topic may benefit from future work that considers users' preconceptions and other factors that influence how users develop and enact their strategies.
    Finally, among the set of strategies we see, there is evidence that the techniques are in alignment with earlier work.
    We found participants' initial strategies were markedly similar to the groupings found by Zhao et al.~\cite{zhao_supporting_2018}.
    In a similar data hand-off task with an \rev{interaction history-like} provenance representation, they identified five typical strategies.
    Their strategies ``random access,'' ``tracing from origin,'' and ``starting over'' closely align with our \textit{random access}, \textit{reviewing origin}, and \textit{starting over} categories, \rev{respectively}.
    The key difference is that we have combined their ``naive browsing,'' and ``hubs and bridges'' categories into one group of \textit{keyword browsing}.
    While they described how some participants adjusted and shifted their strategies, we did not see as many transitions in \rev{our shorter} 30-minute analysis session, likely due to the limited time participants had to complete the task.
    Yet, the definitions they used to describe the various interaction strategies and techniques were in alignment with the set of strategies we observed our participants employ.
    Since their set of categories was also based on traditional, non-collaborative sensemaking strategies, our work further reinforces the idea that hand-off strategies are similar to other data analysis and sensemaking techniques in other settings~\cite{Kang_Gorg_Stasko_2009}.
    Drawing on work for creative collaboration, one way we may learn more from these analysis tasks would be to observe strategies employed \rev{in more creative scenarios and the} user interactions longitudinally~\cite{Gonzales_Fiesler_Bruckman_2015}.
    We also see further evidence in support of the work by Sacha et al.~\cite{sacha_role_2016}.
    They identified how skeptical users will only begin verifying another's work if they see anomalies or have hypotheses about where mistakes were made.
    Those who used a \textit{Reviewing the Origin} strategy were often intrigued by some aspect of the prior analysis and sought to resolve these needs for evidence by reviewing many of the same documents already reviewed. 
    This led to higher ratios of overlap and less independence from the prior analysis.
    \rev{With similarly identified strategies to prior work~\cite{zhao_supporting_2018}, our findings further reinforce that a handful of common strategies exist for different analysis tasks that seem to be based on a user's situated approach.}

\subsection{Limitations}

\rev{Our study of how users pick up an analysis with provenance information from prior collaborators was based on a single analysis scenario and targeted participants with limited data analysis experience.
To build further knowledge on the topic, following research is needed with additional data sets and with participants with varying backgrounds from different collaborative data analysis communities like intelligence operators, medical teams, and academic researchers.}
\rev{Studies could also consider how differences in participants' abilities, problem-solving aptitude, or preferences for particular analytic strategies might influence different approaches or interaction patterns.}

\rev{A challenge with working with open-ended sensemaking tasks is that a user's conclusions may not fit within certain bounds of available conclusions and captured data metrics.
We see this in capturing participant conclusions.}
We set out to capture participant accuracy, error correction, findings made, amount of detail, and depth of relationships but did not have the fidelity in the task complexity nor measures to capture meaningful differences.
In future work, more thoughtful care and constraints should be taken to more formally compare these factors and the influence of provenance.

\rev{In this work, the researchers describe the tool and scenario to participants.
Due to the natural variation in speech, there are potential confounds introduced based on different inflections being interpreted by different users.
While all participants were read the same statements, future work could better control for these effects by using a prerecorded procedure or expanding to additional datasets with different contexts to help generalize findings.}

While our work introduced the comparison of two prototypical techniques used in the literature, there are also more ways of representing past work and their influence on user performance ought to be explored.
for example, some work focuses on generating textual summaries~\cite{Willett_Ginosar_Steinitz_Hartmann_Agrawala_2013,Kandel_Paepcke_Hellerstein_Heer_2011}, while others show branching timelines~\cite{Borland_Wang_Zhang_Shrestha_Gotz_2020,Walker2013Extensible,dunne_graphtrail_2012,walch2020lightguider} that communicate how analysis adapts, transforms, and evolves.
Still, others use graphs and networks as a way of providing concept maps~\cite{zhao_supporting_2018,zhao2013interactive,lung2012inflo}, while others design comics as a technique for summarizing segments of time~\cite{bach2018design,hong2018visual}.
Questions remain not only in the representation provenance should take, but also at what level of detail the provenance should be maintained.
The comparison among how these techniques influence user performance as well as variations in the level of detail ought to be explored in the future.

\section{Conclusion}
In this paper, we examine the effects of provenance representations on future investigator behaviors.
In an open-ended textual data analysis task users were given different kinds of providence information visualizations and asked to pick up the analysis.
We examine the downstream effects of two prototypical provenance representations for collaborative sensemaking (i.e., Coverage and History).
Like the findings of Kang et al.~\cite{Kang_Gorg_Stasko_2009}, while provenance representations do not appear to have significant impacts on user strategy, it does suggest that there are benefits of providing a succinct representation of provenance to help users pick up where other users left off.
While both representations take users time to situate their understanding, we see evidence that the History representation takes more time to comprehend.
Because it does not simplify the prior analysis as well as the Coverage representation, the History representation appears to introduce an additional sensemaking task for participants.
This appears to be especially true when the purpose of provenance is to collaboratively communicate~\cite{Ragan_Endert_Sanyal_Chen_2016}.
It would be interesting to see how these prototypical representations behave when different analysis tasks are evaluated.
Our results imply that provenance summaries that reduce the complexity of the analysis will be beneficial in hand-off analysis scenarios. 
To prevent the overwhelm associated with information overload, perhaps dynamic, user-defined levels of detail in provenance would shed additional light on design guidelines for the analytic provenance community.
These results contribute to the refinement of design guidelines for provenance representations and further emphasize the need for provenance summary techniques.

\acknowledgments{

The authors wish to thank the participants for their involvement in this study. This work was supported in part by the DARPA Perceptually-enabled Task Guidance (PTG) Program under contract number HR00112220005.
This manuscript has been authored by UT-Battelle, LLC under Contract No. DE-AC05-00OR22725 with the U.S. Department of Energy. The United States Government retains and the publisher, by accepting the article for publication, acknowledges that the United States Government retains a non-exclusive, paid-up, irrevocable, world-wide license to publish or reproduce the published form of this manuscript, or allow others to do so, for United States Government purposes. The Department of Energy will provide public access to these results of federally sponsored research in accordance with the DOE Public Access Plan (http://energy.gov/downloads/doe-public-access-plan).
}

\bibliographystyle{abbrv-doi}

\bibliography{main}
\end{document}